\documentclass[conference]{IEEEtran}
\IEEEoverridecommandlockouts

\usepackage{cite}
\usepackage{amsmath,amssymb,amsfonts}
\usepackage{algorithmic}
\usepackage{graphicx}
\usepackage{textcomp}
\usepackage{xcolor}

\usepackage{booktabs}
\usepackage{color}
\usepackage{url}
\usepackage{array}
\usepackage{tabularx}
\usepackage{threeparttable}
\usepackage{adjustbox} 
\usepackage{ragged2e}  
\usepackage[hidelinks]{hyperref} 
\usepackage{xurl} 

\usepackage[T1]{fontenc}
\usepackage{seqsplit}        
\newcommand{\code}[1]{\texttt{\seqsplit{#1}}} 
\newcolumntype{L}[1]{>{\RaggedRight\arraybackslash}p{#1}} 
\newcolumntype{C}[1]{>{\centering\arraybackslash}p{#1}}   
\newcolumntype{Y}{>{\RaggedRight\arraybackslash}X}

\begin{document}
\title{MASCOT: Analyzing Malware Evolution Through A Well-Curated Source Code Dataset}

\author{
\IEEEauthorblockN{Bojing Li\textsuperscript{\textsection},
                  Duo Zhong\textsuperscript{\textsection},
                  Dharani Nadendla,
                  Gabriel Terceros,
                  Prajna Bhandary,
                  Raguvir S,
                  Charles Nicholas\textsuperscript{\dag}
\IEEEauthorblockA{\textit{Department of Computer Science and Electrical Engineering}\\
Baltimore, Maryland, USA\\
Email: \{ji18978, duoz1, dharann1, gtercer1,prajnab1,nv25812,nicholas\}@umbc.edu}
}
\thanks{\textsuperscript{\textsection}Bojing Li and Duo Zhong contributed equally to this work.}
\thanks{\textsuperscript{\dag}Corresponding author: Charles Nicholas (nicholas@umbc.edu).}%
}


\maketitle


\begin{abstract}

In recent years, the explosion of malware and extensive code reuse have formed complex evolutionary connections among malware specimens. The rapid pace of development makes it challenging for existing studies to characterize recent evolutionary trends. In addition, intuitive tools to untangle these intricate connections between malware specimens or categories are urgently needed. This paper introduces a manually-reviewed malware source code dataset containing 6032 specimens. Building on and extending current research from a software engineering perspective, we systematically evaluate the scale, development costs, code quality, as well as security and dependencies of modern malware. We further introduce a multi-view genealogy analysis to clarify malware connections: at an overall view, this analysis quantifies the strength and direction of connections among specimens and categories; at a detailed view, it traces the evolutionary histories of individual specimens. Experimental results indicate that, despite persistent shortcomings in code quality, malware specimens exhibit an increasing complexity and standardization, in step with the development of mainstream software engineering practices. Meanwhile, our genealogy analysis intuitively reveals lineage expansion and evolution driven by code reuse, providing new evidence and tools for understanding the formation and evolution of the malware ecosystem.

\end{abstract}
\section{Introduction}

With the rapid development of information technology and large language models, malware has experienced a surge in recent years, exhibiting strong connections among categories and specimens, as well as high code reuse rates~\cite{calleja2018malsource}.  
In the past 12 months, more than 107 million new malicious or potentially unwanted applications were detected~\cite{av-atlas-malware, stationx-malware-stats}. Many of these malware specimens are variants of previously known malware, which indicates the prevalence of code reuse and family-oriented evolution.

The surge in malware and the high correlation between specimens in the same family make malware evolution analysis an increasingly significant area of research. 

However, the difficulty of collecting, reviewing, and labeling has resulted in a scarcity of source code datasets~\cite{rokon2020sourcefinder}. Existing datasets lack human curation, reliable labels, and timestamps.
Meanwhile, systematic software engineering studies of malware source code evolution date back to 2017~\cite{calleja2018malsource}; the proposed trends and metrics may no longer align with today’s rapidly evolving malware ecosystem.

Current malware genealogy studies primarily focus on binaries and emphasize macro-level clustering and visualizations~\cite{he2019fast,cozzi2020tangled}. However, compared with source code, disassembly and decompilation are prone to ambiguity, inaccuracy, and distortion~\cite{cao2024evaluating}. Accurate unpacking and disassembly are even recognized as major challenges~\cite{cozzi2020tangled}. Therefore, malware genealogy on source code enables more fine-grained and reliable analyses of malware evolution.

This paper presents a well-curated malware source code dataset we call MASCOT, which stands for Malware Source Code Open Treasury. The MASCOT dataset has 6032 Windows-based malware specimens collected from GitHub, making it (we believe) the largest human-reviewed malware source code dataset available. Using MASCOT, we have conducted two studies to analyze malware evolution from the perspectives of software engineering and genealogy.
First, we refer to and expand upon previous research~\cite{calleja2018malsource}, which analyzes malware evolution based on software engineering metrics. Second, we design and implement malware genealogies based on source code reuse.
The corresponding fine-grained, compatible visualization framework can also be applied to genealogies based on disassembled or decompiled code.  Furthermore, our genealogy-based analysis provides an intuitive presentation of the evolutionary process of malware specimens, version by version.

From a software engineering perspective, we conduct experiments in four dimensions to comprehensively measure the evolutionary trends of malware: a) Scale: Source Lines Of Code (SLOC), files, and Function Points (FPs). b) Development costs: effort, time, and team size. c) Code quality: comment-to-code ratio,  cyclomatic complexity, and execution paths. d) Security and dependency: Common Weakness Enumeration (CWE), system calls, and APIs.

From a genealogical perspective, we construct genealogies based on code reuse. Our visualization tool provides both an overall and a detailed view for analyzing global trends and specific evolutionary histories. In the overall view, we present the evolutionary relationships within and between different categories of malware. In the detailed view, we analyze the specific evolutionary history of individual specimens. Also, we present function-level code reuse and function tags to indicate functionalities of both source (reused) and derivative (reusing) functions.

The main contributions of this paper are as follows:
\begin{itemize}

\item We present what is, to our knowledge, the largest publicly available, manually reviewed malware source code dataset. For the main languages used in this dataset (Python, C++, C, C\#, and Assembly), each specimen is annotated with eight standardized labels: Fully Undetectable (FUD), family, behavior, vulnerability, class, file, pack, and unknown. Some specimens, however, lack specific labels due to limitations of labeling tools.

\item From the perspective of software engineering, we systematically analyzed the temporal evolution of malware across four dimensions: scale, development cost, code quality, as well as security and dependency.

\item From the perspective of genealogy, we build and visualize malware genealogies based on source code reuse. Malware evolution across categories and specific specimens can be tracked intuitively from an overall view and a detailed view.

\end{itemize}

\section{Related work}

Other researchers have created malware source code datasets.  For example, in 2017, Calleja et al. presented the MalSource dataset, which contains 456 samples from GitHub and underground forums, 
spanning the years 1975–2016~\cite{calleja2018malsource}.
Rokon et al. proposed a framework for identifying malware code repositories from public platforms, constructing a dataset of 7504 samples with 89\% precision and 86\% recall~\cite{rokon2020sourcefinder}.
VX-Underground and similar platforms also provide large collections of malware source code specimens with preliminary categories such as platform, language, and type~\cite{vxunderground}.

The MASCOT dataset differs from existing malware source code datasets in three aspects: a) Labels: For each specimen written in a major programming language, we provide eight labels. These labels are obtained by parsing VirusTotal reports, a mainstream and trusted annotation method. b)Human review: Based on project comparison feature on GitHub, we filter out trivial fork specimens such as renaming variables, changing parameters, or modifying description files. We believe that filtering out these trivially modified malware specimens can eliminate spurious correlations between samples. c) Metadata: All malware specimens are collected from GitHub with complete development histories and programming language information. For timestamps, we record the earliest development date for original projects and the fork creation date for forked specimens.

Genealogy analysis is a significant perspective to study malware evolution. Cozzi et al. clustered 93,000 IoT malware samples based on binary similarity and constructed genealogy graphs. Through case studies, they analyzed cross-family evolution and code reuse, and later cross-validated their observations using IoT source code~\cite{cozzi2020tangled}. Haq et al. proposed a pipeline comprising unpacking, disassembly, version identification, and lineage construction, which can accurately generate lineage graphs even for wild specimens (which may be packed)~\cite{haq2018malware}.

According to previous research, the primary challenge for constructing Windows-based malware genealogy is controlling errors introduced during unpacking, disassembly, and deobfuscation. The main difference between this paper and existing studies is that malware genealogy based on source code largely avoids such inaccuracies. In addition, our visualization framework provides fine-grained views for in-depth analysis.

\section{Dataset}
\subsection{Data Collection}

\begin{figure}
    \centering
    \includegraphics[scale=1]{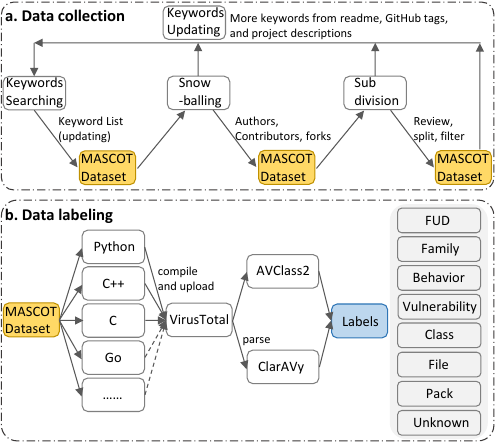}
    \caption{The workflow of data collection and data labeling}
    \label{fig:dataset}
\end{figure}

Building a malware source code dataset is a complex process that relies on manual review. Since malware authors rarely disclose their source code, we can only collect a small part of source code specimens on public platforms or underground forums. To collect standard metadata and improve collection efficiency, we collect and review malware specimens only on GitHub. The complete workflow is illustrated in Figure~\ref{fig:dataset}a.

\begin{itemize}

\item Keyword searching: We collect the first batch of keywords according to news, reports, famous malware families, malware-related vocabularies, synonyms, and abbreviations. Although the first batch of keywords is not comprehensive enough to search for all types of malware specimens, it is continuously expanded during the following processes.
We also use a multilingual version of each keyword to cover non-English malware on GitHub. In this step, we read each repository’s description and ReadMe files to confirm it’s a malicious specimen, and verify that it contains actual source code rather than advertising content.

\item Snowballing: Based on the collected specimens, we consider the evolutionary characteristics and the high code-reuse rate to assume: (a) A malware author (contributor) may publish additional malware. (b) Forks of a given malware often contain variants. For each collected specimen, we analyze GitHub repositories from the same authors (contributors) and explore forks to find variant versions. By using GitHub fork comparisons, we filter out forks that lack essential updates, such as those that only edit Readme files or change parameters.

\item Project Subdivision: Some GitHub repositories are malware datasets, but serve multiple purposes, such as distribution or advertisement. We manually review each specimen in these datasets to remove unnecessary files and retain only those containing actual source code, thereby maintaining the quality of MASCOT.

\item Keyword updating: During manual review, we skim descriptions or ReadMe files to confirm malware source code and explore new keywords. As the number of reviewed specimens increases, the keyword list becomes more comprehensive and can cover a wider range of malware-related scenarios. In turn, the expanded list helps us to find more malware specimens in subsequent iterations.

\end{itemize}

\subsection{Basic information on MASCOT}

\begin{figure}[t]
    \centering
    \includegraphics[scale=0.68]{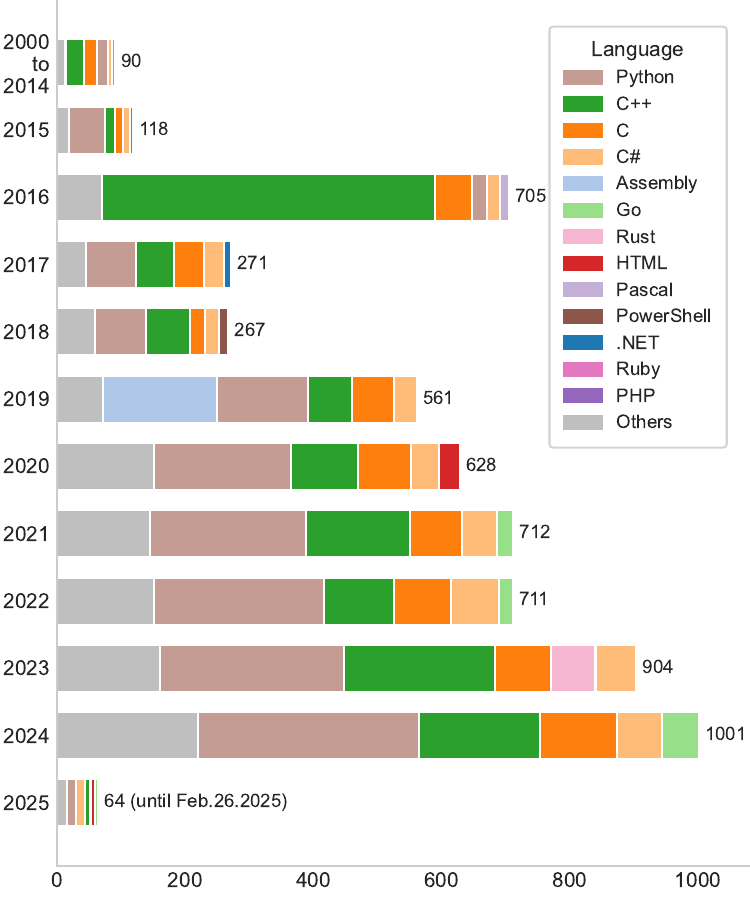}
    \caption{Top-5 malware languages every year}
    \label{fig:language-malware figure}
\end{figure}

After several iterations of the above processes, we built MASCOT, a Windows-based malware source code dataset containing 6032 specimens with dates ranging from November 6, 2000, to February 26, 2025.
As shown in Table~\ref{tab:programming-languages}, there are 32 different programming languages used in the MASCOT dataset.
Although Java is widely used in general software~\cite{index2024languages}, malware authors seem to prefer scripting languages such as Python.
Figure~\ref{fig:language-malware figure} presents the number of malware specimens in MASCOT each year along with the top five programming languages. 
In general, the number of malware specimens has increased every year except 2017 (further discussed in Section 5). 
Python, C++, and C have been the most popular languages among malware authors in most years.
Since 2021, newer languages such as Go have enjoyed increasing levels of adoption.

\begin{table}[t]
  \centering
  \caption{Number of specimens in different programming languages}
  \label{tab:programming-languages}
  \renewcommand{\arraystretch}{1.1}
  \setlength{\tabcolsep}{2.5pt} 
  \begin{tabularx}{\columnwidth}{@{}*{5}{>{\centering\arraybackslash}X}@{}}
    \toprule
    Python & C++ & C & C\# & Assembly \\
    1765   & 1568 & 691 & 449   & 241 \\ \midrule
    Go     & HTML & Powershell & Rust & Others \\
    185    & 169  & 163        & 124  & 677   \\
    \bottomrule
  \end{tabularx}
\end{table}

\begin{table}[b]
\centering
\caption{Label counts after parsing VirusTotal reports}
\renewcommand{\arraystretch}{1.1} 
\begin{tabularx}{0.987\linewidth}{lXXXXXXXX}
\toprule
  & File & Fam & Vuln & Beh & Class & Pack & FUD & Unk \\
\midrule
Cnt & 783 & 111 & 30 & 269 & 368 & 5 & 115 & 195 \\
Pct & 67.4\% & 9.6\% & 2.6\% & 23.2\% & 31.7\% & 0.4\% & 9.9\% & 16.8\% \\
\bottomrule
\end{tabularx}
\label{tab:number-of-labels}
\end{table}


\subsection{Labeling}

Malware source-code datasets are underutilized partly due to missing or inaccurate labels, such as family, category, and behavior. 
As shown in Figure~\ref{fig:dataset}b, we manually reviewed and compiled each project according to its ReadMe. Then we uploaded the executables to VirusTotal~\cite{virustotal_home} and parsed the resulting reports to obtain eight labels by using AVClass2~\cite{sebastian2020avclass2} and ClarAVy~\cite{joyce2025claravy}. Before manually reviewing, we upload the folder structure and ReadMe to GPT\mbox{-}4o for compilation suggestions. In this process, we believe the worst-case scenario for such use of an LLM is that it cannot compile or obtain labels, but it won't modify source code or produce incorrect labels. Therefore, its benefits outweigh the potential risks.

Labeling malware is time-consuming. We first compile and label the samples written in the five most prevalent programming languages in the dataset, which are Python, C++, C, C\#, and Assembly. These labels support the subsequent experiments in this paper and lay the foundation for future research. Other labels in MASCOT may be released in the future.

As cross-language analyses of software engineering metrics and code reuse detection are still in their early stages, we chose C/C++ specimens for subsequent experiments because of their long history and widespread use in malware analysis.
For 2259 C/C++ specimens in MASCOT, 627 already included executables; we compiled 766 of the remaining specimens. Due to VirusTotal constraints, we uploaded and parsed 1161 of 1393 executables and obtained their labels.

As shown in Table~\ref{tab:number-of-labels}, parsing VirusTotal reports does not provide all eight labels (file, family, vulnerability, behavior, class, pack, FUD, and unknown) for each specimen. For example, among 1161 uploaded specimens, only 368 successfully obtained the Class label, accounting for 31.7\%. We noted, with some concern, that even though parsing VirusTotal reports has become the mainstream labeling approach, only 9.6\% of the specimens obtained a family label. Furthermore, 9.9\% of the specimens were identified as FUD, meaning they were determined to be malicious based on their Readme but were not detected by VirusTotal.

\section{Malware Evolution From a Software-Engineering Perspective}

To extend the timeline of malicious source code, we combine the MalSource dataset~(1975-2016)~\cite{calleja2018malsource} with MASCOT and adopt the following analytical methods from a software engineering perspective: We adopt the three evaluation aspects used in MalSource (scale, development costs, and code quality) in the first three parts of this section. These consistent metrics allow us to observe malware evolution systematically from 1975 to 2025. 
Then, we add a fourth part, security and dependency (system calls, APIs, and CWEs), to capture the recent trends in malware code security and external dependencies.

\subsection{Scale}
\begin{itemize}

\item{Source Lines of Code:} SLOC measures a project from the perspective of code volume. The physical SLOC that we are using is to count the number of code lines and exclude comments or blank lines. 

\item{Number of Files:} The number of files measures a project from the perspective of file organization and provides a valuable complement to SLOC. It counts all source files and excludes those without any programming logic behavior in C/C++ projects, such as .json and .xml files.

\item{Function Points:} FPs measure the size of a project from the perspective of functionality. FPs are language-independent and analyze how a program behaves with users. For example, inputs, outputs, software interaction with users, files, databases, and how the software connects to other systems are all examples of function points.
\end{itemize}

\subsection{Development Costs}
\begin{itemize}

\item{Effort:} Effort refers to the total amount of human work estimated for developing a project, measured (for example) in person-months. The COCOMO model~\cite{boehm1984software} calculates the effort metric according to Eq.~(\ref{equ:effort}), where $a$ and $b$ are varying constants that depend on the type of projects.

\begin{equation}
    \text{Effort (PM)} = a \times (\text{SLOC}/1000)^b
    \label{equ:effort}
\end{equation}

\item{Time (months):} The development time measures the duration required to complete a project. The COCOMO model assumes the whole project is distributed across small teams and further proposes Eq.~(\ref{equ:time}) to measure development time. 

\begin{equation}
    \text{DevTime} = 2.5 \times (\text{Effort})^{0.38}
    \label{equ:time}
\end{equation}

\item {Team Size:} This metric indicates the number of full-time developers needed to build a project within a given estimated development time. We estimate this value by dividing total effort by development time in the COCOMO model.
\end{itemize}

\subsection{Code Quality}
\begin{itemize}

\item{Comment-to-Code Ratio:} This metric compares the comments and source code to evaluate readability and further measure the code quality of a project. However, malware authors sometimes modify comments to avoid detection and analysis~\cite{stockley_trojan_source_2021}.

\item{Cyclomatic Complexity:} Cyclomatic complexity measures the number of independent paths in a function. It counts the minimum number of ways a function can be tested based on decision points (if, else, while). Malware often exhibits a high cyclomatic complexity in order to discourage interpretation and analysis.

\item{Execution Paths:} Execution Paths measure the number of ways a program can run based on logical statements. Each new decision point increases the execution paths of the program. This metric helps to study the behavior of a malware specimen. 

\end{itemize}

\begin{figure}[t]
    \centering
    \includegraphics[width=\columnwidth,height=0.4\textheight]{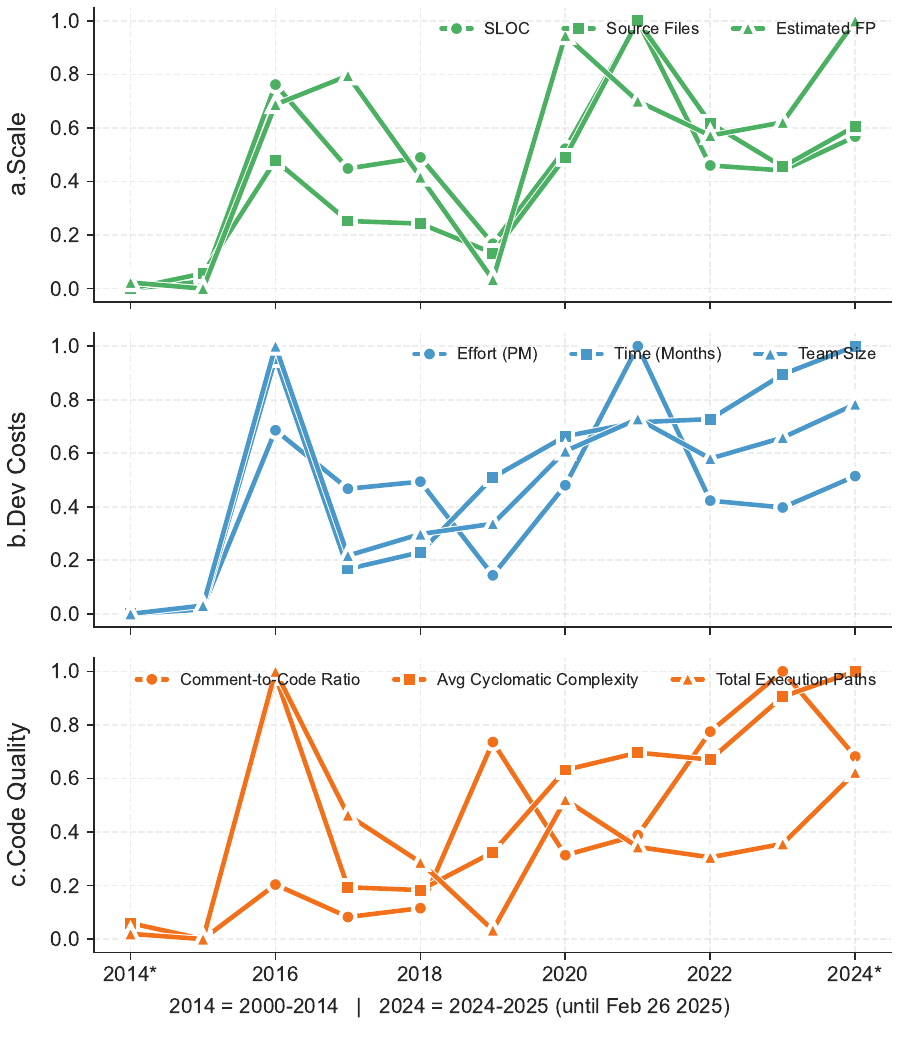}
    \caption{Temporal Trends of Malware Metrics (Normalized)}
    \label{fig:malware-temporal-figure}
\end{figure}

\subsection{Insights on Scale, Dev Costs, and Code Quality}

Figure~\ref{fig:malware-temporal-figure} and Table~\ref{tab:dharani-summary} show the results of scale, development costs, and code quality. We have two observations: 
First, malware development has become standardized and increasingly complex over time, with some fluctuations.
The trends for metrics in each dimension are similar, which indicates consistent evaluation conclusions from different perspectives. The overall trends suggest that malware development is getting standardized and systematic, as evidenced by the increase in scale, development cost, and code quality. Some outliers, such as 2016, suggest potential large-scale human-driven malware development, as shown in Fig.~\ref {fig:language-malware figure} and reports~\cite{korolov2016} (which will be further analyzed in Section 5A).

Second, compared to early specimens, modern malware is lightweight but of high quality. As Table~\ref{tab:dharani-summary} illustrates, modern samples generally exhibit smaller code sizes and lower development costs while maintaining superior code quality metrics. This trend aligns with some contemporary developments: scripting languages are getting popular; more functions are provided in packages; and complex code logic is being transferred to cloud platforms. Malware development is following modern software engineering trends: functionality isn’t shrinking; instead, high-quality code encapsulates complex malicious behaviors.

\subsection{Security and Dependency}

\begin{table}[t]
  \centering
  \caption{Statistics Comparison Between MalSource and MASCOT}
  \label{tab:dharani-summary}
  \renewcommand{\arraystretch}{1.2}
  \begin{tabularx}{\columnwidth}{L{1.65cm} L{1cm} Y Y Y}
    \toprule
    \textbf{Scale}  & Size & SLOC(avg) & Files(avg) & FP(avg) \\
    \midrule
    MalSource2017  & 456 & 6567 & 17 & 66 \\
    MASCOT2025    & 6032 & 5721 & 14 & 57 \\
    \bottomrule
  \end{tabularx}


\begin{tabularx}{\columnwidth}{L{1.65cm} L{1cm} Y Y Y}
    \midrule
    \textbf{Dev Costs}  & Size & Effort(avg) & Time(avg) & Team(avg) \\
    \midrule
    MalSource2017  & 456 & 19.37 & 4.38 & 1.41 \\
    MASCOT2025    & 6032 & 17.47 & 3.61 & 1.11 \\
    \bottomrule
  \end{tabularx}

\begin{tabularx}{\columnwidth}{L{1.65cm} L{1cm} Y Y Y}
    \midrule
    \textbf{Code Quality}  & Size & CR(avg) & CC(avg) & EP(avg) \\
    \midrule
    MalSource2017  & 456 & 17.69 & 2.06 & 311.55 \\
    MASCOT2025    & 2025 & 18.09 & 3.73 & 502.09 \\
    \bottomrule
  \end{tabularx}

\vspace{0.4em}
{\scriptsize
\parbox{\columnwidth}{
  \textbf{Notes.}\;
  CR = Comment-to-Code Ratio,\quad
  CC = Cyclomatic Complexity,\\
  EP = Execution Paths,\quad
  FP = Function Points
}}
\end{table}


We now present brief descriptions of other attributes present in MASCOT.

\begin{itemize}

\item{CWEs:} CWE is a category system for hardware and software weaknesses and vulnerabilities~\cite{cwe2025}. 
Analyzing CWEs in malicious source code is informative for malware evolution: 
First, CWEs indicate the quality and maturity of malicious specimens. 
Second, some CWE entries (e.g., CWE-506: Embedded Malicious Code) are direct indicators of malicious behavior.
Third, malicious specimens could reuse the same defective code and inherit the same CWEs. 
Last but not least, exploiting vulnerabilities in malware is a novel and realistic defensive strategy that has recently been reported~\cite{quinn2020_emocrash,eriksson2008_securityguru}.

\item{System Calls:} System calls are programmatic ways to request services and interact with the operating system~\cite{Wikipedia_System_call}. Malware execution frequently triggers system calls, making their frequency and sequence critical features for malware dynamic analysis~\cite{kolosnjaji2016deep,canali2012quantitative,xiao2019android}. Recent research further treats sequences such as system calls as low-level languages that can be modeled with natural language techniques for malware analysis~\cite{oak2019malware}.

\item{APIs:} APIs are interfaces used between computers or programs to provide functionality and services. Compared to system calls, APIs are more numerous and comprehensive, not limited to kernel space, but also able to interact with other programs to implement complex functionality. Understanding APIs in malware source code helps us to gain deeper insights into the logical structure of malicious code. Therefore, API sequences are a common and significant feature that has been widely used for malware analysis~\cite{ki2015novel,cui2023api2vec,li2022novel}.

\end{itemize}

\vspace{1.5mm}
\noindent\textbf{Insights on Security and Dependency}
\vspace{0.2\baselineskip}

\begin{table}[t]
  \centering
  \caption{Top 10 CWEs in malware source code}
  \label{tab:cwe-rank}
  \setlength{\tabcolsep}{4pt}      
  \renewcommand{\arraystretch}{1.1}

  \begin{tabularx}{\columnwidth}{@{} L{13mm} L{8mm} Y @{}}
    \toprule
    CWE & Count & Description \\
    \midrule
    
CWE-398  & 2360 & Code Quality \\
CWE-561  & 2048 & Dead Code \\
CWE-563  & 1764 & Assignment to Variable without Use \\
CWE-476  & 954 & NULL Pointer Dereference \\
CWE-571  & 894 & Expression is Always True \\
CWE-457  & 837 & Use of Uninitialized Variable \\
CWE-570  & 783 & Expression is Always False \\
CWE-686  & 775 & Function Call With Incorrect Argument Type \\
CWE-788  & 615 & Access of Memory Location After End of Buffer \\
CWE-401  & 568 & Missing Release of Memory after Effective Lifetime \\

    \bottomrule
  \end{tabularx}
\end{table}
We use Cppcheck~\cite{cppcheck} to detect CWEs in C/C++ specimens in Mascot and Malsource datasets~\cite{calleja2018malsource}. As shown in Table~\ref{tab:cwe-rank}, we list the top 10 CWEs in MASCOT and MalSource, written in C/C++. 
Even as malware development becomes more standardized and complex, CWEs reveal widespread, long-term code defects across multiple dimensions of malware source code. 
In particular, CWE-398, CWE-561, and CWE-563 are among the most prevalent defects in malware source code, providing a potential possibility of proposing new malware indicators by analyzing these CWEs.

\begin{table}[b]
  \centering
  \caption{Mydoom Case: Code Reuse with Vulnerability Inheritance}
  \label{tab:mydoom-cwe}
  \renewcommand{\arraystretch}{1.2}
  \begin{tabular}{p{1.4cm} >{\centering\arraybackslash}p{3.2cm} >{\centering\arraybackslash}p{3.0cm}}
    \toprule
    & Top-20 specimens that reuse Mydoom & All C/C++ specimens \\
    \midrule
    CWE467 & 13/20 = 65\% & 161/2259 = 7.1\% \\
    CWE476 & 14/20 = 70\% & 692/2259 = 30.6\% \\
    Both   & 10/20 = 50\% & 137/2259 = 6.0\% \\
    \bottomrule
  \end{tabular}
\end{table} 

Moreover, CWEs show strong potential for evolution analysis and clustering. Taking malware \code{Mydoom} as a case study: As shown in Table~\ref{tab:mydoom-cwe}, among the top 20 specimens that reused its code, CWE-467 appeared in 65\%, whereas it appeared in only 7.1\% of all C/C++ specimens; CWE-476 appeared in 70\% of \code{Mydoom} derivative specimens, compared to 30.6\%. Specimens containing both CWE-467 and CWE-476 accounted for 50\% in the Mydoom derivative set, versus only 6.0\% overall.

These statistically significant differences~(p<0.001) indicate that during malware evolution, code reuse not only transfers functionality but also inherits vulnerabilities. 
Given that malware authors don't always maintain their source code with care, and that vulnerability detection on executables is often more accurate than on source code, these results introduce a vulnerability-based perspective for tracing malware evolution and genealogy.

Syscalls and APIs: We collect, parse, and deduplicate data from GitHub~\cite{j00ru_windows_syscalls} and Windows metadata~\cite{xlang2025, win32metadata2025, win32metadata2024preview} to obtain two lists: 2379 system calls and 18072 APIs. Then we conduct keyword searching to detect system calls and APIs in each C/C++ specimen. The resulting observations are as follows:

(a) System call and API usage across periods of time reveal malware evolution from multiple perspectives. For system calls, 
Table~\ref{tab:syscall-count} shows the top three most frequently used system calls during different time periods. Before 2020, these frequently used system calls could be divided into two categories: one was to steal system information from victims, such as \texttt{NtQueryInformationProcess}~\cite{behling2025ntquery}. Another concerns preparations for further malicious activities. For example, \texttt{NtOpenFile} is used to open related malicious programs~\cite{symantec2019ransomware}, and \texttt{NtAllocateVirtualMemory} is for allocating virtual memory~\cite{behling2025ntprotect}, a first step in the process of code injection.
However, more straightforward forms of malicious behavior have become popular in recent years. \texttt{NtRaiseHardError} is often used in malicious operations that directly trigger the infamous blue screen on a Windows system without administrator privileges~\cite{AgnivaMaity_NtRaiseHardError-Example_2024,Reddit_CProgramming_BSOD_2022,NtDoc_NtRaiseHardError_2025}.

\begin{table}[t]
\centering
\caption{Top Three System Calls in Malware Source Code by Period}
\renewcommand{\arraystretch}{1.3}
\begin{tabularx}{\columnwidth}{@{} L{7mm} Y L{20mm} Y @{}}
\toprule
Period & Top1\_SysCall(cnt) & Top2\_SysCall(cnt) & Top3\_SysCall(cnt) \\
\midrule

1976\newline-2010 & NtQueryInformation\newline Process~(8) & NtQuerySystem\newline Information~(6) & NtCreateFile~(4) \\

2011\newline-2015 & NtQueryInformation\newline Process~(11) & NtQuerySystem\newline Information~(11) & NtOpenThread~(10) \\

2016\newline-2020 & NtQuerySystem\newline Information~(182) & NtOpenFile~(31) & NtAllocateVirtual\newline Memory~(28) \\

2021\newline-2025 & NtRaiseHardError\newline (84) & NtQuerySystem\newline Information~(81) & NtWriteVirtual\newline Memory~(47) \\
\bottomrule
\end{tabularx}
\label{tab:syscall-count}
\end{table}

For APIs, as shown in Table~\ref{tab:top3-apis-by-period}, from 1975 to 1995, socket primitives such as \code{bind}, \code{connect}, and \code{getsockname} suggest that early specimens established connections directly with low-level calls (sometimes through their own listeners). From 1996 to 2005, \code{CreateThread} and \code{CloseHandle} indicate the frequent usage of Windows multithreading and handle-based resource management. From 2006 to 2015, the commonly used \code{sleep} function was designed to evade antivirus software by intentionally delaying the onset of malicious activity. \code{GetLastError} indicates that error handling was widespread to make malware more robust. From 2016 to 2020, the combination of \code{send}, \code{sleep}, and \code{CloseHandle} appears to be a frequent feature of Command and Control (C2) logic. Malware authors can launch multiple attacks and avoid detection through C2 connections. From 2021 to 2025, \code{GetProcAddress} became popular for enabling dynamic API resolution, reducing static analysis features, and making the malware's purpose harder to discern.

(b) System calls and APIs act as direct indicators and stealthy indicators, respectively. For system calls, many of them have been identified as direct indicators in reports. \code{NtAllocateVirtualMemory} is widely used for remote memory allocation and advanced code injection~\cite{malforge_ntallocatevirtualmemory_2025}. \code{NtWriteVirtualMemory} is used for quieter and harder-to-detect code injection, helping to dodge userland hooks~\cite{behling_ntwritevirtualmemory_2025}. \code{NtQueryObject} is used for anti-reversing techniques to hamper or prevent debugging~\cite{oscar404_antidebuggingtechniques_2025}. \code{NtUnmapViewOfSection} can be used to clear memory space and hijack execution~\cite{behling_ntunmapviewofsection_2025}. \texttt{NtCreateSymbolicLinkObject} can manipulate symbolic links to bypass security programs and escalate privileges~\cite{ibm_xforce_10979}.

For APIs, some may be useful as stealthy indicators of malware. Some of the commonly used APIs among malware are rarely found in benign software. For example, \code{sleep} was intended to cause a program or thread to pause for a specific time. But in malware, it is often used to implement time-based evasion by adding delays in the code~\cite{malapi_sleep}, making that function a fairly strong malware indicator. \code{GetProcAddress} was originally used to obtain the memory address of a function within a DLL. Still, malware authors often exploit it to load malicious payloads and evade antivirus software through indirect calls~\cite{cocomelonc_malware_av_vm_evasion_part16_2023}.  Most of the APIs in this table were collected by MalAPI~\cite{malapi_homepage_mrd0x_2021}, and they have also been confirmed by other security researchers to be associated with malicious behavior. For example, \code{GetLastError} is frequently used by malware authors as part of anti-debugging measures~\cite{guttman_common_antidebugging_2017}.

\section{Malware Genealogy}

\begin{table}[t]
  \centering
  \caption{Top three APIs in malware source code by period}
  \label{tab:top3-apis-by-period}
  \setlength{\tabcolsep}{4pt}      
  \renewcommand{\arraystretch}{1.1}

  \begin{tabularx}{\columnwidth}{@{} L{12.2mm} L{21.7mm} L{21.7mm} Y@{}}
    \toprule
    Period & Top1\_API(cnt) & Top2\_API(cnt) & Top3\_API(cnt) \\
    \midrule

1976-1995 & bind (1) & connect (1) & getsockname (1) \\

1996-2000 & CloseHandle (5) & CreateThread (4) & GetFileSize (4) \\

2001-2005 & CloseHandle (17) & sleep (17) & connect (16) \\

2006-2010 & sleep (29) & GetLastError (26) & CloseHandle (24) \\

2011-2015 & sleep (48) & CloseHandle (41) & GetLastError (36) \\

2016-2020 & sleep (781) & CloseHandle (714) & send (710) \\

2021-2025 & CloseHandle (507) & sleep (418) & GetProcAddress~(332)\\

\bottomrule
\end{tabularx}
\end{table} 

Code reuse is pervasive in malware and creates a detectable lineage both within and across categories. Malware genealogy seeks to identify and characterize these intra-category and inter-category connections to reveal patterns of evolution.
Genealogy analysis at the source code level is, we believe, the most direct and accurate way to recover those connections, because source code can reflect the authors' intention without introducing decompilation noise. 
To our knowledge, the MASCOT dataset is the first malware source code dataset to provide a large-scale collection of malware specimens with C/C++ labels, making malware source code genealogy analysis possible with that data.

\subsection{Data Preparation}

We perform source code level malware genealogy for all C/C++ specimens in MASCOT and MalSource. 
We split each specimen into functions according to the following concerns: On the one hand, shared libraries and common headers could introduce spurious similarity. On the other hand, developers typically use functions as the smallest units for code reuse and updates. 
We then compare functions across specimen pairs and aggregate the results to build malware genealogy and characterize evolutionary trends.

\begin{figure}[t]
    \centering
    \includegraphics[scale=1]{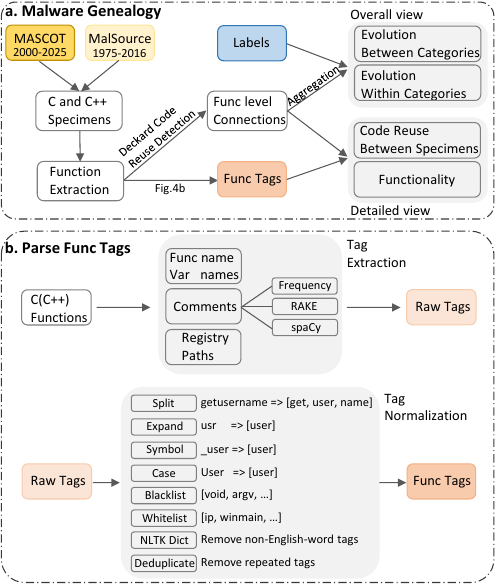}
    \caption{The workflow of malware genealogy and function tags}
    \label{fig:genealogy}
\end{figure}

\subsection{Function Tags}

A key question in function-level reuse detection is, which functionalities are actually reused? To address this, we design a function-tag-parsing pipeline, as shown in Figure~\ref{fig:genealogy}b. When analyzing function-level code reuse across malware specimens, function tags indicate the functionality of both reused and reusing functions, providing a clear understanding of the malware’s fine-grained evolutionary process.

As shown in Figure~\ref{fig:genealogy}b, there are two stages in parsing function tags: tag extraction and tag normalization. 
In tag extraction, we extract developer notes of functionality from function names, variable names, and comments.
We also detect Windows registry and file paths because they are commonly used for modifying system settings. 
Function and variable names are typically concise English words or abbreviations, while comments are sentences. We parse comments into raw tags according to three methods: frequency, RAKE~\cite{rose2010automatic}, and spaCy~\cite{spacy}. 

For tag normalization, we employ the following steps: (1) Split raw tags written in camelCase or joined by symbols into words. (2) Expand each abbreviation/alias where applicable. (3) Clean symbols and lowercase all raw tags. (4) Filter out non-English words using NLTK tools~\cite{bird2006nltk} while retaining tags in the whitelist and removing tags in the blacklist. (5) Deduplicate the processed tags. Table~\ref{tab:func_tag} shows that function tags and function names clarify the functionality of source code.

\subsection{Code Reuse Detection}

\begin{table}[b]
\centering
\renewcommand{\arraystretch}{1.25}
\setlength{\tabcolsep}{4pt}
\caption{Examples of Funcion Tags and Functionality}
\label{tab:func_tag}
\begin{tabularx}{\columnwidth}{
  @{}>{\raggedright\arraybackslash}p{1.2cm}
     >{\raggedright\arraybackslash}p{1.4cm}
     >{\raggedright\arraybackslash}p{2cm}
     >{\raggedright\arraybackslash}X@{}}
\toprule
Speci & Func & Tags & Manual Review \\
\midrule

\parbox[t]{1.9cm}{Botnet\\\_Win32\\.Mimail} & grab\_start & grab; start; found; print; window; send & Monitor window titles; once a specific target matches, grab the data and send it. \\

\midrule

\parbox[t]{1.9cm}{wke\\\_sdbot\\\_i3s} & cpuspeed & speed; start; something; cycle; count & Measure the cycle count during 1 second to estimate the CPU frequency. \\

\midrule

\parbox[t]{1.9cm}{chntpw\\-master\\\_Turing} & ex\_next\_v & next; count; parse; data & Enumerate items in the custom structure and parse their contents. \\

\bottomrule
\end{tabularx}
\end{table}


In code reuse detection, we apply the Deckard method~\cite{jiang2007deckard}. The Deckard method is based on measuring the structural similarities between abstract syntax trees. It has been widely adopted in code reuse detection due to its accuracy, scalability, and cross-language adaptability. Since it focuses on structural information rather than lexical letters, it can detect code reuse even in cases of different variable names.

\begin{figure}[t]
    \centering
    \includegraphics[scale=1]{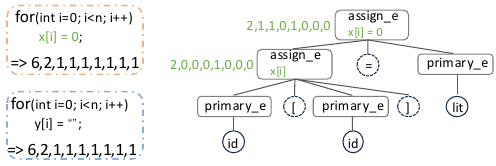}
    \caption{Deckard parses code into an AST and summarizes it with vectors}
    \label{fig:deckard}
\end{figure}

As shown in Figure~\ref{fig:deckard}, the Deckard method starts by parsing the source code into an abstract syntax tree. Then it aggregates the structural information of each subtree in a bottom-up fashion. This structural information includes the frequencies of each type of nodes (\code{id}, \code{lit}, \code{assign\_e}, \code{incr\_e}, \code{array\_e}, \code{cond\_e}, \code{expr\_s}, \code{decl}, and \code{for\_s}) to form a fixed-length feature vector. Finally, Deckard applies locality sensitive hashing for clustering and thereby identifies code reuse.

\subsection{Genealogy}

\begin{figure*}[t]
    \centering
    \includegraphics[scale=1]{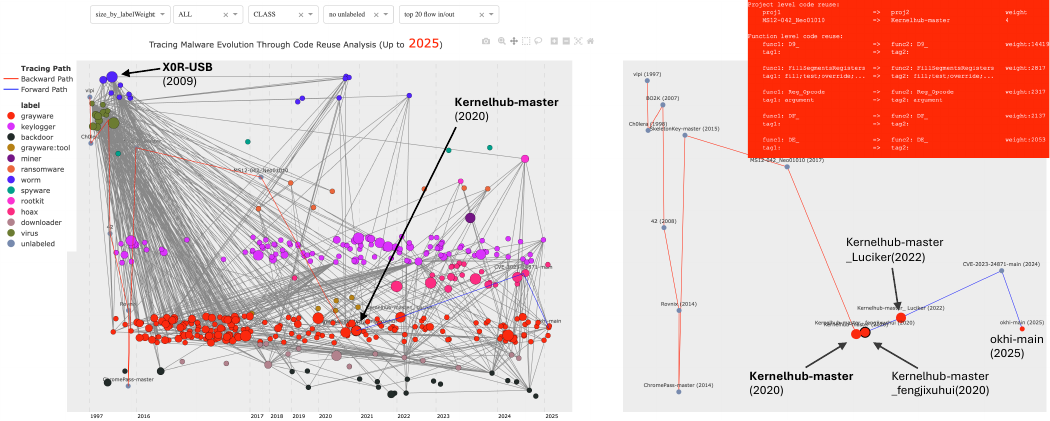}
    \caption{Example of a genealogy under the class label, where the left side is an overall view and the right side is a detailed view. }
    \label{fig:gene-vis}
\end{figure*}

Malware genealogy analysis aims to map evolutionary connections between malware categories (families) and within categories over time.
Malware authors often refer to existing malicious source code and modify it to fit their own projects. This process of reusing malicious source code is like the flow of genes, recurring across and within malware categories (families) and gradually forming a complex malware genealogy. 
Combining results from code-reuse detection with temporal information (the MASCOT dataset includes timestamps), we establish evolutionary connections between malware specimens. To our knowledge, this is the first large-scale analysis of malware genealogy at the source code level.

When observing the evolutionary genealogy of malware, we analyze from both an overall view and a detailed view.
From the overall view (category-centric), we ask questions such as: which families' specimens are most frequently reused by others? Which families reuse the least existing code? Which families reuse the most from others? To answer these questions, we aggregate function-level code reuse results back to the specimen level. The weight between two specimens is defined as the sum of code reuse weights across all their functions. Finally, we group specimens by label to derive evolutionary relationships and corresponding strength (weights) at the category (family) level.
From the detailed view (specimen-centric), we focus on the complete evolutionary history of a single malicious specimen: Which prior specimen is its primary source of reused code, that is, its direct parent? Which subsequent specimen (its direct child) reuses its code? The parent–child relations based on a given specimen form a local lineage and temporal evolutionary history. Moreover, according to function-level code reuse detection, we can therefore know "whether the code reuse happened" but also precisely determine "which functions were reused" and even "what is the functionality of reused functions". By combining the function tags with the results of function-level code reuse detection, we reconstruct a fine-grained evolutionary process for a single malware specimen.

\subsection{Virualization}

Current malware lineage analyses typically rely on clustering-based visualizations that distinguish families or other labels by color, while additional information needs manual annotation. We present a new genealogy visualization framework with the following advantages: a) Multi-label perspectives: Our visualization can switch lineage views according to different labels, such as family, behavior, and class. b) Temporal information: X-axis encodes calendar year; it also provides year-by-year animation for evolutionary changes. c) Overall and detailed views: The framework provides both an overall view and a detailed view to analyze trends and the evolutionary history of a specific specimen. d) Compatibility: The visualization framework is open-source and compatible for malware lineage research based on multiple features rather than source code.

\section{Evolutionary Insights From Malware Genealogy}

\subsection{Insights From An Overall View of Malware Genealogy}

According to the parsed results from AVClass and ClarAVy, each specimen has eight labels: class, behavior, file, FUD, vulnerability, pack, and unknown. Each label provides an overall genealogy for analyzing different categories of malware and a detailed perspective on the evolutionary history of a single specimen. 
We use the "class" label as an example to illustrate our observation; other genealogy visualizations are available~\cite{MASCOT}.

As shown in Figure~\ref{fig:gene-vis}, there are 12 categories represented by different colors, all under the class label: greyware, keylogger, backdoor, grayware tool, miner, ransomware, worm, spyware, rootkit, hoax, downloader, and virus. To analyze the temporal evolution of different categories, we notice temporal shifts in category prevalence: viruses were more prevalent around 1997-2007, while hoaxes have been more popular in recent years. Meanwhile, keyloggers and grayware have maintained substantial numbers over the past two decades, undergoing continuous maintenance and evolution. Among all keywords used to collect the MASCOT dataset, Keylogger (including keystroke) retrieved the largest number of specimens (29.33\%). This suggests that keyloggers may be the most widespread and longest-enduring malware category.

Regarding grayware, we observed an abnormal peak in 2016, which is higher than in other years. This batch of grayware exhibits substantial code reuse, particularly with the virus and worm clusters from around 1997 to 2007. Considering reports that 27\% of malware variants were created in 2015~\cite{korolov2016}, we infer that malware authors extensively referenced previous virus/worm source code to create new variants. This batch of emerging malware was classified as grayware, and its source code was released in 2016.

As shown in Figure~\ref{fig:gene-vis}, malware authors developed groups of virus and worm malware around 2000. These early specimens have been continuously referenced and modified over more than two decades and have propagated across multiple categories, including grayware, keyloggers, and backdoors, becoming a core source of cross-category reuse. 
A representative case is~\code{X0R-USB} (2009), which exhibits worm, backdoor, and DDoS capabilities. Its function~\code{WinMainCRTStartup} was reused by several botnet malware in 2016 as a backdoor entry point. 
Its DDoS function was reused by~\code{remote\_hacker\_probe} in 2021. And its thread-listening code was reused in 2010 by botnet specimens for the email monitoring module~\code{email\_check2}. 
A sobering observation is that some specimens reuse malicious functions from~\code{X0R-USB} without even renaming, yet remain undetected by more than seventy AV engines on VirusTotal.
Overall, code reuse is not confined to the same family or category, indicating that authors reuse code from classic malware across different types.

\subsection{Insights From A Detailed View of Malware Genealogy}

The detailed view provides us with the specific evolution trajectory of an individual malware specimen. By clicking on a node on the overall view, the detailed view reveals the history of this specimen in the pervasive code-reuse connections. The detailed view centers on an individual specimen and presents its complete evolutionary paths: which existing malware specimen it reused, and which subsequent specimen reused it. Meanwhile, the detailed view provides function-level code reuse connections: which specific functions were reused, and what are the potential purposes of these functions (function tags). 

As shown in Fig.~\ref{fig:gene-vis}, when we click on the specimen \code{Kernelhub-master}, an evolution history spanning 1997 to 2025 is displayed, connecting 13 malware specimens. On this path, red and blue represent ancestors and descendants, while the thickness of the path represents the weight of code reuse. The clicked specimen was reused by user \code{fengjixuhui} in 2020 and by user \code{Luciker} in 2022, with both keeping the original name of this malware. This phenomenon provides strong evidence for the evolutionary path constructed by our genealogy-based method. 

There are two additional points worth noting in the detailed view. First, according to the evolutionary path of Kernelhub-master, only 3 out of 13 specimens have resolved the class label by VirusTotal, AVClass2, and ClarAVy. According to Table~\ref{tab:number-of-labels}, the class label has the second parsing rate among all eight labels. This indicates that antivirus engines still have significant challenges in identifying malware classes, families, behaviors, and other labels.
Second, according to the function-level code reuse information in Fig.~\ref{fig:gene-vis}, \code{Kernelhub-master} reuses at least five functions from its parent specimen \code{MS12-042\_Neo01010}. Among the reused functions, these two specimens not only have identical function names and tags, but also exhibit tags that cannot be parsed due to short function names, variable names, and comments. These evidences indicate that \code{Kernelhub-master} directly borrowed a substantial portion of code from \code{MS12-042\_Neo01010}. However, \code{MS12-042\_Neo01010} was not identified by antivirus engines as containing the class label. This phenomenon warns us that during code reuse, even without substantial modifications to the source code, antivirus engines could be misled during label detection.

\section{Discussion}

We discuss the observations and limitations of our work in this section. We conduct experiments from both a software engineering and a genealogy perspective and have some observations. For example, the sudden surge of malware source code specimens in 2016 may be related to the outbreak of malware variants in 2015. 
Compared to earlier malware, modern malware specimens tend to be smaller in scale, have lower development costs, but higher code quality.

We further observed and demonstrated that during code reuse, the vulnerabilities of malware itself are inherited. Given that malware authors typically focus more on functionality than robustness of their own code, correlation analysis based on “vulnerability inheritance” offers a novel perspective for malware lineage research.
In addition, System calls and APIs serve as straightforward and stealthy indicators of malicious behavior, respectively.  The prevalent APIs in different periods align with the development of software engineering. From a genealogy perspective, our visualizations demonstrate the extensive code reuse from early malware specimens. We also provide a detailed view to show the evolutionary history of individual specimens, which is a powerful tool for analyzing specific specimens.

Nevertheless, this study has limitations. First, the GitHub platform returns 100 pages of results for every keyword search, thus most of our specimens are from 2014 to 2025. Second, even though parsing through VirusTotal is one of the most trustworthy labeling methods,  a significant number of samples lack labels, particularly malware family labels. Third, since we plan to conduct code reuse experiments at the function level, comparing functions rather than specimens individually incurs extremely high computational costs. Although Deckard is an efficient code reuse detection tool, the false positives it introduces can interfere with evolutionary analysis.

All the specimens in MASCOT are from GitHub, which presents advantages and introduces bias. On one hand, GitHub records the full development history of each project, which enables us to have accurate timestamps and filter out forks without substantial updates. On the other hand, malware authors also disseminate code from underground forums and other channels. Thus, generalizing observations derived from all GitHub specimens to entire malware ecosystem inevitablly introduces bias.

Data availability. The disarmed dataset will be released on IEEE DataPort. Related statistical files and visualization code is available on the GitHub repository for this paper~\cite{MASCOT}.

\section{Conclusion}

In this paper, we present a well-curated malware source code dataset and analyze malware evolution from a software engineering perspective and a genealogy perspective. Our MASCOT dataset contains 6032 human-reviewed malicious specimens and provides eight labels for most of specimens. From a software engineering perspective, we analyze the scale, development costs, code quality, as well as security and dependency of malware. The result indicates that malware development is evolving toward a lightweight yet high-quality direction, in line with mainstream software engineering trends. We also present a novel perspective for analyzing malware genealogy through the inheritance of vulnerabilities.
From a genealogy perspective, our lineage tools make that evolution measurable. We demonstrate the connections between and within different malware categories and provide a detailed view for analyzing the evolutionary history of individual specimens. Case studies and a potential human-driven large-scale malware development in 2015-2016 have also been analyzed according to genealogy.
We hope that both MASCOT dataset and the genealogy tool can serve as a foundation for reproducible malware-evolution studies in the future.

\bibliographystyle{Style/ieee.bst}
\bibliography{references}
\end{document}